# Physical Computing at the Data Processing Inequality Limit

Yuhang Zheng[1]†, Yang Zhao[1]†, Xiuting Zou[1], Chunyu Zhao[1], Zhiyi Yu[1], Zechen Li[1], Jiaxing Wu[1], Shaofu Xu[1]*, Weiwen Zou[1,2]*

[1]State Key Laboratory of Photonics and Communications, Intelligent Microwave Lightwave Integration Innovation Center (imLic), School of Integrated Circuits (School of Information Science and Electronic Engineering), Shanghai Jiao Tong University; Shanghai 200240, China.

[2]State Key Laboratory of Micro-nano Engineering Science, School of Integrated Circuits (School of Information Science and Electronic Engineering), Shanghai Jiao Tong University; Shanghai 200240, China.

*Corresponding author: Weiwen Zou (wzou@sjtu.edu.cn), Shaofu Xu (s.xu@sjtu.edu.cn)

†These authors contributed equally to this work.

**Abstract:** Wave-physics-based intelligent sensing has driven multidisciplinary applications from smart industries to decision-making systems. Traditional sensing paradigms transform physical waveforms into human-understandable intermediate representations through preprocessing. Such transformations inherently cause information loss owing to data processing inequality (DPI). Here, we established a theoretical framework for physical computing at the DPI upper limit. Physical computing avoids information loss during preprocessing by directly extracting information from physical waveforms, achieving the theoretical maximum of accessible information as determined by the DPI. Furthermore, physical computing comprehensively utilizes multiple dimensions of physical waveforms, thereby enhancing the upper limit of information capture capability. Electromagnetic sensing experiments have demonstrated that physical computing can achieve 100% sensing accuracy, substantially outperforming traditional sensing paradigms. The proposed theoretical framework of physical computing offers a promising path towards enhancing the information-capture capability of next-generation intelligent sensing systems.

Advancements in intelligent sensing and decision-making technology have substantially increased the autonomy of smart agents in radar detection[1–3], autonomous driving[4], smart homes[5], and healthcare[6]. The nature of sensing via wave physics is to capture information from waveforms that interact with the environment and its targets[7]. Enhancing information-capture capability is the foundation of intelligent sensing and nearly all smart agents, as shown in Fig. 1a[8–12]. Contemporary intelligent sensing systems follow humanoid sensing paradigms to capture information while suppressing noise and interference[13,14]. The received physical waveforms, such as electromagnetic waves, acoustic waves, and light waves, are converted into intermediate representations that are easily understood by humans, typically images[15–17], point clouds[18,19], spectra[20,21], and other visual forms. Subsequently, feature extraction and inference were performed based on these intermediate representations to achieve the final decision (Fig. 1b).

According to information theory, mutual information quantifies the amount of information on target features contained in physical waveforms[22]. In classification models, mutual information defines the minimum probability of error that can be achieved[23]. The data processing inequality (DPI) states that any information processing flow conforming to the Markov chain $X \to Y \to Z$ must satisfy: $I(X;Z) \leq I(X;Y)$, where the mutual information about the source $X$ contained in data $Y$ represents the theoretical upper limit for all subsequent processing stages[24]. DPI plays an instructive role in fields such as wireless communication[25–27], quantum communication[28,29], and machine learning[30,31]. In sensing and decision-making, when physical waveforms are converted into intermediate representations for human visualization, subtle but crucial information for final decisions may be discarded or blurred[32]. Therefore, implementing physical computing directly with physical waveforms is expected to overcome the limitations inherent in traditional sensing paradigms, thereby enabling higher decision-making accuracy. Recently, physical computing has been demonstrated using acoustic and electromagnetic prototypes to demonstrate their low-power consumption and low-latency properties[33–37]. However, physical computing still lacks theoretical insights from the perspective of information theory and a framework for strengthening the information capture capability to outperform traditional humanoid sensing paradigms.

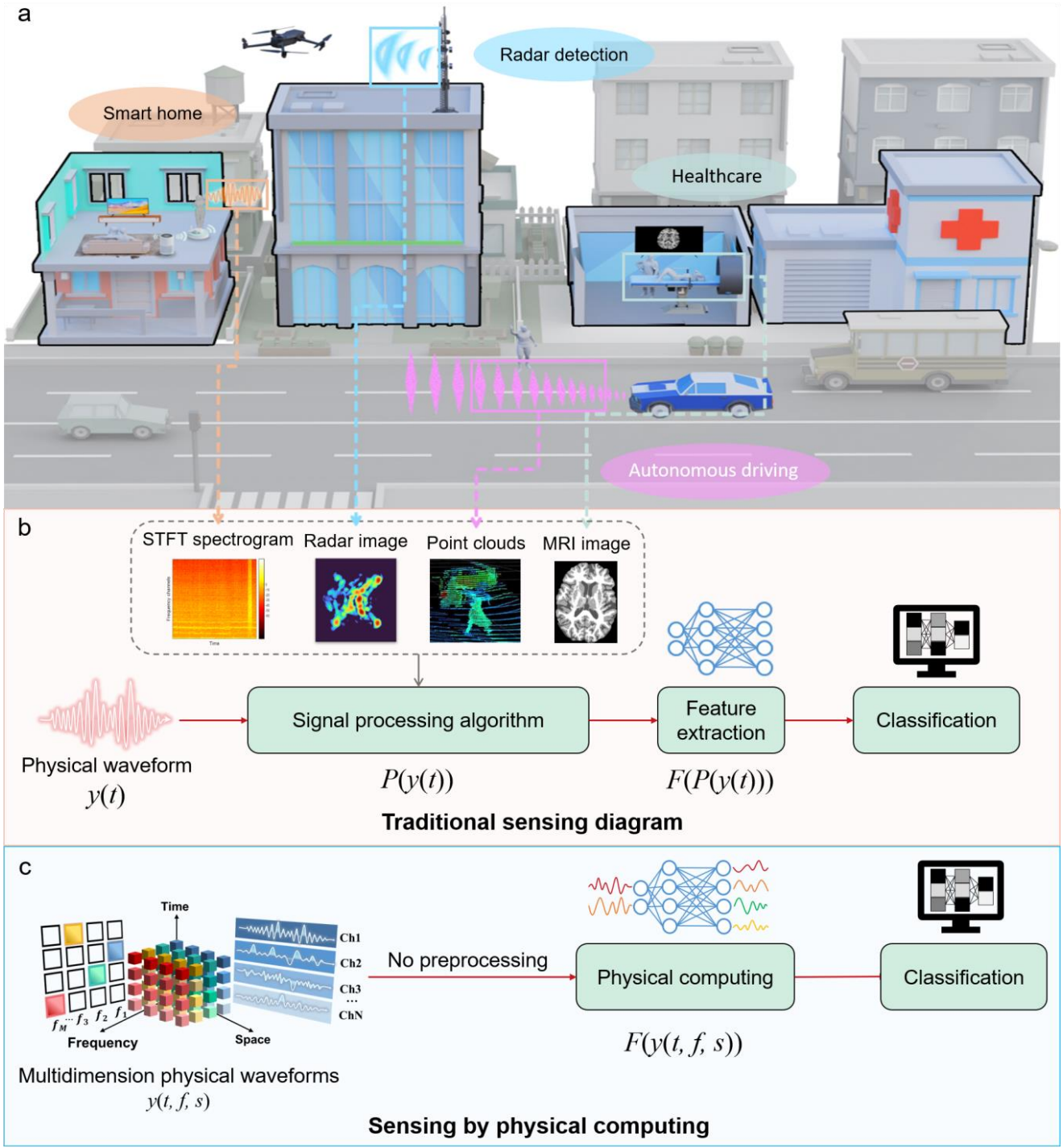

**Fig. 1 | Overview of physical computing approaches. a,** Intelligent sensing and decision-making are widely applied in smart homes, radar detection, autonomous driving, and healthcare. Systems perform functions such as voice recognition, target recognition, road perception and disease diagnosis based on acoustic, electromagnetic, light wave physics and magnetic. **b,** Traditional sensing paradigms. The physical waveform $y(t)$ is first converted into intermediate results $P(y(t))$ (such as spectra, images, point clouds, etc.) through signal processing algorithms. These intermediate representations are classified through feature extraction and inference to achieve intelligent sensing. **c,** Sensing by physical computing (this work). Multiple dimensions of physical waveforms $y(t, f, s)$ are utilized to increase the information. Physical computing is directly performed without extra preprocessing, and the extracted information $F(y(t, f, s))$ is classified to achieve high-accuracy sensing.

In this study, we propose a theoretical framework of physical computing for high-performance intelligent sensing, as shown in Fig. 1c. We present theoretical and experimental evidence confirming the effectiveness of DPI, demonstrating that physical computing can preserve more information than traditional sensing paradigms. The received physical waveforms are directly processed to extract critical information without human-understandable preprocessing, thereby achieving the upper limit of the DPI. Based on this, we propose a multi-dimensional method for wave physics to increase the upper bound of mutual information defined by the DPI. This method effectively captures critical information from physical waveforms. We evaluated the performance of physical computing in a target recognition task and achieved 100% accuracy—significantly higher than traditional sensing paradigms.

## Results
### Information retention by physical computing

To demonstrate that physical computing can achieve the DPI limit, we conducted a target recognition experiment of electromagnetic sensing as proof of concept. More details are provided in the Materials and Methods section. Consider the Markov chain $h \to y \to P(y)$, where $h$ is the target impulse response (TIR), $y$ is the received physical waveforms and $P(y)$ is the intermediate representation after signal processing[38]. According to the DPI, $I(y;h) \geq I(P(y);h)$. We found that information loss occurred during the imaging process (Supplementary note S1). The inequality can therefore be rewritten as $I(y;h) > I(P(y);h)$. Thus, we contend that the information obtained through physical computing is greater than that derived from radar images, enabling higher recognition accuracy through the classifier.

The physical waveforms obtaining architecture and recognition module we use for physical computing is shown in Fig. 2a. We employed different methods to process electromagnetic waveforms for radar target recognition and conducted ablation studies. Figure 2b displays the different targets used in our experiments. The high-resolution range profiles (HRRPs) of the targets are shown in Fig. 2c, respectively. As evident from the results, different target shapes exhibited varying numbers of scattering points and different scattering intensities in the HPPR (for instance, drone and J16 showed fewer scattering points). Two-dimensional inverse synthetic aperture radar (ISAR) images shown in Fig.2d clearly reflect the shape characteristics and detailed features of the targets (such as the four supporting legs of the drone and the canard wings of the J20). We used a linear frequency-modulated (LFM) signal as the transmit signal to compare the effectiveness of target recognition by physical computing versus different imaging-based methods, the processing of which adheres to DPI principles.

In physical computing, at the input, the physical echo waveform passed through two one-dimensional convolutional layers with 3 kernels, resulting in 4 channels of feature representations followed by max-pooling. Finally, the output was flattened and passed through two fully connected layers to produce the final predictions. Cross-entropy was employed to measure the discrepancy between the predicted outputs and the ground truth labels. Throughout training and validation, the CNN achieved a convergence accuracy of approximately 1 and a loss near 0, indicating no overfitting and validating the choice of kernel design and loss function (Fig. S3a). To demonstrate the effectiveness of information extraction, we applied t-stochastic neighbor embedding (t-SNE) as a dimensionality reduction tool to examine extracted features. Figures 2e-h depict the confusion matrices and t-SNR feature results of different processing methods for physical waveforms (Supplementary note S2 provides the detailed training process): (1) applying physical computing to extract information directly from electromagnetic waveforms for recognition, (2) performing pulse compression on physical waveform to obtain target HRRP, then extracting features from both amplitude and phase components for recognition, (3) after pulse compression of physical waveform to obtain target HRRP, then extracting features from only the amplitude component for recognition, and (4) using the BP algorithm to construct ISAR image from physical waveform, then extracting features from the radar image for recognition. For methods (1), (2), and (3), the neural network configurations were identical, consisting of two one-dimensional convolutional layers, whereas method (4) employed a neural network configuration with three convolutional layers[39]. The scale of the network parameters in method (4) was larger than that in methods (1), (2), and (3). Comparing methods (1) and (2), their recognition accuracies were very close (89% vs. 88%), which is consistent with DPI because the pulse compression process is reversible and no mutual information is lost. Meanwhile, their t-SNE features were also similar: data from drones and J16 formed clusters with clear boundaries from other classes, but the features of J20 and Plane were similar and difficult to distinguish. Method (3), unlike method (2), discards phase information, resulting in mutual information loss and consequently lower recognition accuracy. The t-SNE features show that after losing phase information, the feature separation between categories becomes smaller, especially as the J20 and Plane features become confused. The confusion matrix shows that even after characteristic mapping through the fully connected layers, accurate classification cannot be achieved. Finally, the results of method (4) demonstrate that even images distinguishable to humans cannot achieve good recognition performance after feature extraction, as features from different classes are difficult to separate. This is due to mutual information loss caused by the integration process in the ISAR imaging algorithms and the final operation of obtaining the intensity images. These experimental findings prove that the direct processing of electromagnetic waveforms by physical computing preserves more mutual information, enabling higher recognition accuracy.

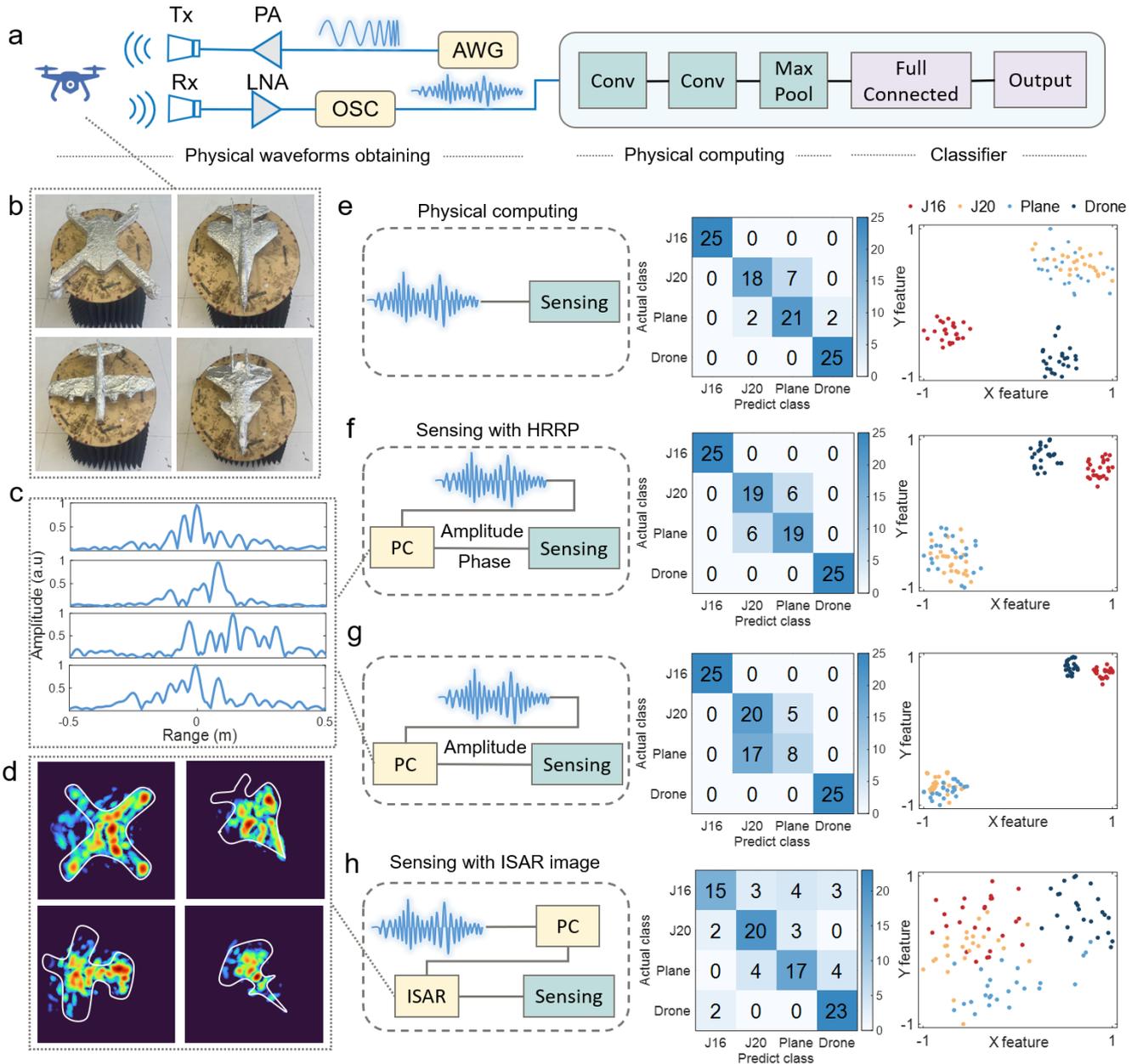

**Fig. 2 | Experimental demonstrations of physical computing achieving the DPI limit in radar target recognition. a,** Brief physical waveforms obtaining architecture and recognition module of physical computing. AWG arbitrary waveform generator, PA power amplifier, Tx transmitter, Rx Receiver, LNA low-noise amplifier, OSC oscilloscope, Conv convolution. **b,** Photographs of the targets "Drone", "J16", "Plane" and "J20". **c,** High-resolution range profiles (HRRPs) of the targets, respectively. **d,** ISAR images of the targets, respectively. **e-h,** Confusion matrices and t-SNE results of different processing methods. **e,** Physical computing of physical waveforms. **f,** After pulse compression of physical waveforms, feature extraction is performed on both the real and imaginary parts of the HRRP. PC pulse compression. **g,** After pulse compression of physical waveforms, feature extraction is performed only on the magnitude of the HRRP. **h,** After ISAR imaging of physical waveforms, feature extraction is performed on ISAR images using a 2D convolutional neural network. ISAR inverse synthetic aperture radar.

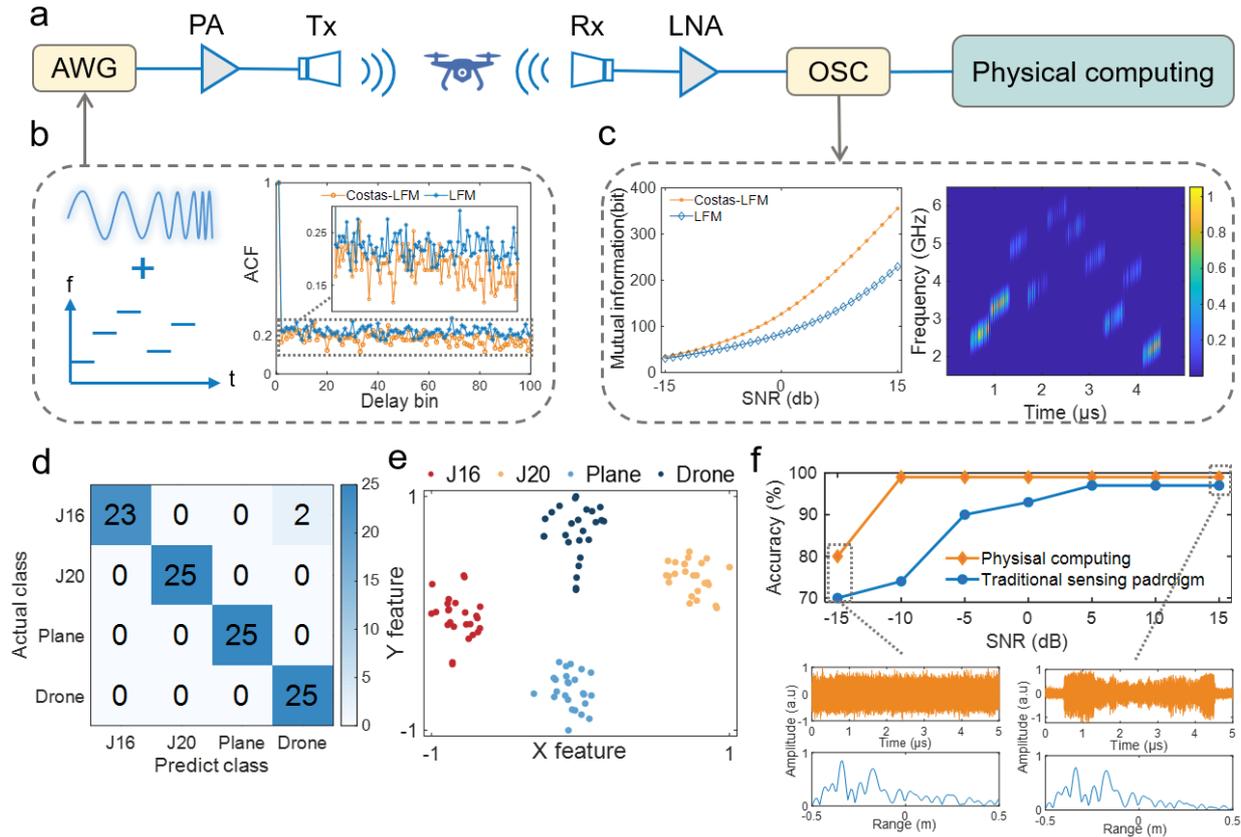

**Fig. 3 | Information enhancement characteristics of Costas-LFM in physical computing. a,** Brief experimental architecture of sensing by Costas-LFM. **b,** Comparison of Autocorrelation functions of Costas-LFM and LFM. **c,** Mutual information from different waveforms under different signal-to-noise ratio (SNR) conditions and the time-frequency image of received Costas-LFM waveform. **d,** Confusion matrices of physically computing Costas-LFM physical waveforms. **e,** t-SNE results for physically computing Costas-LFM physical waveforms. **f,** Comparison of recognition accuracy between physical computing and traditional sensing paradigm under different SNR conditions. Figures on the left and right sides present the physical waveforms and pulse compression results under SNR of -15dB and 15dB, respectively.

**Enhancing frequency dimensions of physical computing**

The DPI indicates that applying physical computing can preserve more information from physical waveforms. By utilizing time-frequency dimension of the physical waveforms, information can be directly enhanced, thereby advancing the information limit of the DPI by physical computing. In radar target recognition, we theoretically prove that when the autocorrelation function of the transmit signal approaches zero, the mutual information increases. A detailed derivation of the formula is provided in the supplementary note S3. Owing to its low cross-correlation and frequency agility, Costas-encoded LFM with carrier frequency modulation is commonly used in anti-jamming radars[40]. Here, we selected it as the transmit signal, enabling full utilization of the temporal and frequency dimensions of the physical waveforms to obtain more mutual information (Fig. 3a). Costas-LFM is derived by applying Costas coding to the carrier frequency of LFM signals, and its non-zero autocorrelation function is lower than that of LFM (Fig. 3b). When Costas-LFM is used as the transmit signal, the time-frequency diagram of the received waveform and the comparison of obtained mutual information are displayed.in Fig. 3c.

To validate that Costas-LFM can carry more mutual information, we performed physical computing for recognition using Costas-LFM physical waveforms. Figures 3d and 3e show the confusion matrix and t-SNE results obtained from physical computing and recognition of single-antenna physical waveforms, demonstrating a significant improvement in recognition performance compared to LFM physical waveforms (89% to 98%). This indicates that the features of each data category were distinctly clustered, with different feature clusters separated from each other. Therefore, the experimental results demonstrate that, compared to LFM physical waveforms, transmitting Costas-LFM enhances the mutual information. By enabling the received physical waveforms to acquire more mutual information, the system can achieve high-precision target recognition. For comparison, we also conducted recognition on the Costas-LFM HRRP, achieving 97% accuracy, which closely matched the performance of physical computing. Detailed results are discussed in the supplementary note S4. Experimental results show that receiving Costas-LFM physical waveforms yields more mutual information than LFM by utilizing time-frequency dimension characteristic, and even after imaging operations cause information loss, the retained mutual information remains sufficient for accurate recognition.

Because the physical computing of physical waveforms lacks a pulse compression step, under low-SNR conditions, the physical waveforms can be submerged in noise. To investigate whether physical computing can effectively extract target information from high intensity noise, we compared its recognition accuracy with that of the traditional sensing paradigm (with pulse compression) under different SNR conditions, as shown in Fig. 3f. Supplementary note S5 provides details of the training process. Under SNR conditions of -15 dB and 15 dB, the physical waveforms and pulse compression results are presented on the lower left and lower right sides, respectively. At -15 dB, the physical waveform is completely submerged in noise, while the pulse compression result remains nearly unaffected. However, across the SNR range from -15 dB to 15 dB, the recognition accuracy of physical computing consistently exceeds that of the traditional sensing paradigm. Moreover, physical computing accuracy remains very high level until the SNR drops below -10 dB. This indicates that even under low-SNR conditions, physical computing can effectively utilize mutual information from noise to achieve higher recognition accuracy, demonstrating its feasibility for engineering applications in real-world environments.

**Enhancing spatial dimensions of physical computing with photonic processor**

Photonics offers unique advantages for physical computing by directly processing physical waveforms in the analog domain[41,42]. Additionally, owing to their broadband capacity and high parallelism[43–46], photonic systems can carry and process spatial-time-frequency dimension of physical waveforms, thereby further enhancing the information upper bound. Therefore, we propose a photonic physical computing processor (PPCP) that implements physical computing in photonic hardware. As illustrated in Fig. 4a, the PPCP utilizes a convolutional neural network architecture and leverages analog photonic circuits for information extraction[47]. According to the principle of an integrated photonic tensor convolution processor, the analog waveforms of Costas-LFM from different angles are modulated onto optical carriers of different wavelengths, and then fed into an optical processing core (OPC) for physical computing[48]. The OPC multiplexes (MUX) optical carriers in single optical waveguides, and an array of delay lines applies consistent temporal delays to input waveforms. At each delay step, the micro-ring weighting banks multiply the input waveforms with convolutional weights. The number of micro-rings in each weighting bank corresponds to the number of wavelengths, and each micro-ring is precisely tuned to modulate transmittance for a specific wavelength. Finally, the optical waveforms are converted to electrical signals using photodetectors (PDs), optical power across all wavelengths is summed, and an electronic power combiner (EPC) accumulates electrical signals from all delay steps. Feature signals are sampled using an oscilloscope (OSC). Nonlinear (ReLU) and MaxPooling (pool size = 8) operations are implemented in the digital domain, and a fully connected layer generates the final recognition results.

In target recognition experiments, transfer learning was adopted to enhance recognition accuracy[34]. The feature signals acquired by the OSC were used to retrain the fully connected layers implemented in the digital domain. The training results are presented in Fig. 4b. The loss functions of both the training and validation sets decreased following similar trends and eventually stabilized, whereas the recognition accuracy consistently increased until convergence. The validation accuracy reached 100%. As shown in Fig.

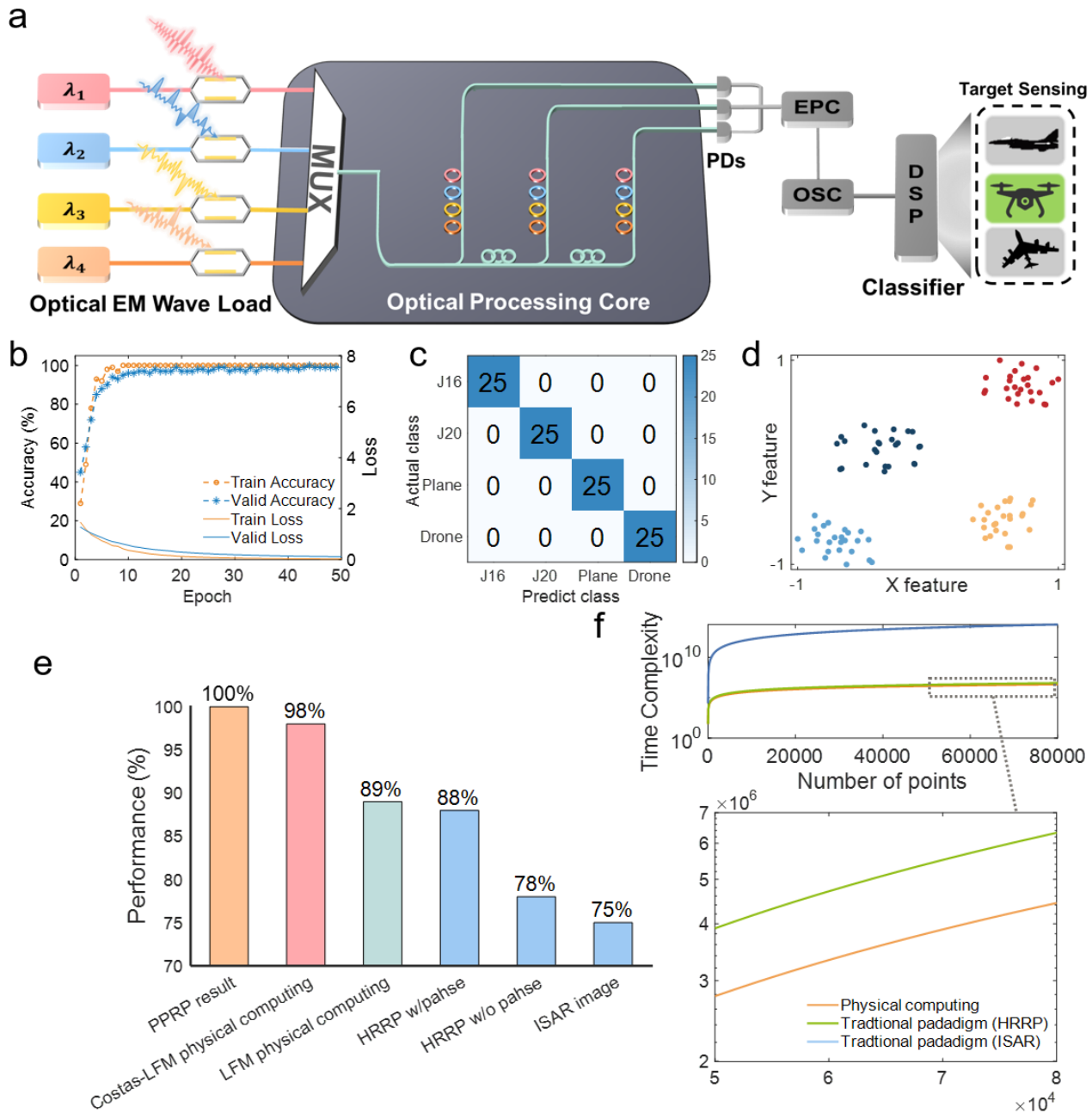

**Fig. 4 | Physical computing with PPCP and the effectiveness of physical computing. a,** Electromagnetic waveforms from different spatial dimensions are modulated into optical carriers with different wavelengths, while the optical processing core performs a convolution-equivalent operation. Multiple optical waveforms are multiplexed; micro-rings of different wavelengths apply weights to different channels, and delay lines perform shift operations equivalent to convolution. PDs convert the optical waveforms to electrical signals, which are then summed via EPC to complete the convolution operation. The OSC captures the signals. DSP implements nonlinear and MaxPooling operations, and the fully connected layer outputs the recognition results. **b,** Loss functions and classification accuracy of transfer learning. **c,** t-SNE results of feature maps. **d,** Confusion matrices of classification. **e,** Comparison of recognition accuracy across different processing methods. "PPCP result" corresponds to the accuracies in Fig. 4b; "Costas-LFM physical computing" corresponds to Fig. 3b; "LFM physical computing," "HRRP w/ phase," "HRRP w/o phase," and the "ISAR image" corresponds to Figs. 2e-h. **f,** Time complexity of physical computing, traditional paradigm by HRRP, and traditional paradigm by ISAR.

4c, even without passing through the fully connected layers, the t-SNE features extracted from each target category tended to cluster together, and clear separations existed between the features of different target categories. The confusion matrix of the recognition results is shown in Fig. 4d, where all categories were correctly classified. These results indicate that the PPCP can effectively extract information for high-accuracy classification, ensuring that the fully connected layers in the digital domain after transfer learning achieve high-precision target recognition.

Figure 4e shows the recognition accuracy of different transmit waveforms and signal processing configurations discussed previously. The LFM physical computing results demonstrate that direct waveform processing through physical computation achieves superior accuracy compared to traditional sensing paradigms. A comparative analysis of PCPP, Costas-LFM physical computing, and LFM physical computing indicates that comprehensive utilization of spatiotemporal-frequency multidimensional physical waveforms enables target discrimination with more information. The trend in recognition accuracy clearly shows that using multidimensional waveforms and physical computing significantly improved performance, validating both the theoretical and experimental effectiveness of the proposed approach.

## Discussion

By applying physical computing to waveforms, our method not only achieves higher sensing accuracy by reaching the DPI limit but also simplifies the sensing processing workflow, resulting in lower latency compared to traditional sensing paradigms. We analyzed the time complexities of three processing methods for radar target recognition: physical computing, HRRP recognition, and ISAR image recognition, as shown in Fig. 4f. Further details are provided in supplementary note S6. ISAR imaging requires the accumulation of multiple physical waveforms, yielding the longest time for target recognition, with a time complexity of $O(N^3)$ or $O(N^2)$. Although the processing method for HRRP recognition requires pulse compression, its time complexity is $O(NlogN)$. By contrast, the physical computation of waveforms processes the signals directly; hence, the time complexity is $O(N)$.

In this study, we propose a physical computing sensing framework that achieves the upper limit of the DPI. Theoretically, we demonstrate that compared to traditional sensing paradigms, physical computing possesses the advantage of directly extracting information from physical waveforms, enabling efficient information processing based on DPI. Furthermore, by comprehensively utilizing multiple dimensions of the physical waveforms, the total information content carried by the waveforms can be effectively increased, thereby raising the DPI upper bound. In our proof-of-concept experiment, a target-recognition task validated the physical computing concept, achieving 100% sensing accuracy—much higher than traditional sensing paradigms. Numerical analyses verified the effectiveness of physical computing and multidimensional utilization of physical waveforms.

In the intelligent era, the trend towards achieving more intelligent sensing tasks using smart agents is inevitable. Our approach establishes a theoretical foundation for directly extracting information from physical waveforms and demonstrates the significant potential of physical computing for intelligent sensing.

## Methods
### Experimental setup

The experimental setup for physical computing is shown in Fig. 2a. In the experiment, the radar transmit signal was generated using an arbitrary waveform generator (AWG, Keysight, M8195A) and passed through a power amplifier (Talent Microwave, TLPA1G18G-40-33). The target was located approximately 3 m away from the transmitting and receiving antennas. In the PPCP experiments, the transceiver antennas for different channels were positioned adjacently, following multi-input multi-output (MIMO) antenna distribution principles and practical configurations[49], with each antenna set distributed in an arc at 36° intervals. After receiving the physical waveforms of the target, a low-noise amplifier (LNA, Connphy, CLC-10M6G-3440S) was used to enhance the SNR. An oscilloscope (OSC, Keysight, DSO-S) was then used to capture the output signals, which were used for physical computing.

The transmitted signal was an LFM signal with a starting frequency $f_0$ of 2 GHz and a bandwidth of 4 GHz. If the transmitted signal is a Costas-LFM signal, it is encoded by the Costas sequence $N=$ [2,4,8,5,10,9,7,3,6,1]. Consequently, the minimum carrier frequency spacing $\Delta f$ is 0.4 GHz, where the carrier frequency for each part of the signal is $f_i = f_0 + N(i) \times \Delta f, i = 1,2,…,10$, and each part of the segment has a bandwidth of 0.4 GHz. The signal has a sampling duration of 5 μs and a pulse duration of 4 μs. The time-frequency image of the received physical waveform is shown in Fig. 2c, illustrating that the carrier of each sub-signal follows the Costas coding pattern.

In the PPCP experiments (Fig. S1), we employed an OPC to realize a two-layer convolution operation with a stride. The four-channel signals were modulated onto optical carriers of different wavelengths (C32, C34, C36, and C38), generated by a four-channel laser source, using Mach-Zehnder modulators (MZM, EOSPACE, AX-0MVS-40-PFA-PFA-LV). For both transmission and reception, the microwave line lengths were precisely equal, and tunable optical delay lines were used to compensate for optical path length variations between the modulator and OPC input ports, ensuring consistent signal time delays. Additional delay lines were employed at the OPC output to adjust time delays for proper implementation of the strided convolution. Subsequently, three photodetectors (PD, CONQUER, KG-PT-10G-SM-FA) performed optical-to-electrical conversion and power accumulation. The signals were then recorded by the OSC and fed into the fully connected layer for classification using a computer.

**CNN setup and training**

The neural network used in the physical computing experiment comprised two convolutional layers and two fully connected layers. The convolutional kernel parameters for both first and second convolutional layers were [input channels, output channels, kernel width] = [4, 4, 3], with a stride of 2, and the activation function was ReLU. The numbers of neurons in the first and second fully connected layers were set to 300 and 4, respectively. The activation functions for these layers were ReLU and sigmoid, respectively. The neural network used for ISAR image recognition comprised three convolutional layers and two fully connected layers. The convolutional kernel parameters of the first convolutional layer were [input channels, output channels, kernel size] = [1, 48, 3 × 3]; for the second layer, [48, 128, 3 × 3]; and for the third layer, [128, 128, 3 × 3]. The activation function for each convolutional layer was ReLU. After each convolutional layer, a 2 × 2 max-pooling layer was applied. The numbers of neurons in the first and second fully connected layers was set to 300 and 4, respectively, with ReLU and sigmoid as their activation functions.

To obtain the training dataset, 50 physical waveforms were collected for each target category, yielding a total of 200 waveforms. One hundred waveforms were randomly selected for training, and the remaining 100 were used for validation. Similarly, for ISAR image recognition, 50 images were obtained for each target category, for a total of 200 images. One hundred images were randomly selected for training, and the remaining 100 were used for validation.

For the PPCP experiment, using the aforementioned models and data, we first trained the neural network and used the convolutional layer parameters as OPC parameters. During transfer learning, only the parameters of the fully connected layers were retrained. One hundred groups of feature maps were selected for training, and the remaining 100 groups were used for validation.

**Acknowledgements**
This work is supported in part by the National Natural Science Foundation of China grant T2225023 (W.Z.) and grant 62205203 (S.X.).

**Author contributions**
The study was conceptualized by Y. Zheng, X.Z. and W.Z. The experiments were conceived by Y. Zheng, C.Z., Z.Y., Z.L and J.W. The theoretical derivation was accomplished by Y. Zheng, X.Z. and C.Z. Data analysis was carried out by Y. Zheng and Y. Zhao. The optical processing core was designed by S.X. All authors contributed to the writing. The project was supervised by W.Z., and S.X.

**Data availability**
All data used in this study are available from the corresponding authors upon reasonable request.

**Conflict of interest**
The authors declare no conflict of interest.

# Physical Computing at the Data Processing Inequality Limit

Yuhang Zheng[1]†, Yang Zhao[1]†, Xiuting Zou[1], Chunyu Zhao[1], Zhiyi Yu[1], Zechen Li[1], Jiaxing Wu[1], Shaofu Xu[1]*, Weiwen Zou[1,2]*

[1]State Key Laboratory of Photonics and Communications, Intelligent Microwave Lightwave Integration Innovation Center (imLic), School of Integrated Circuits (School of Information Science and Electronic Engineering), Shanghai Jiao Tong University; Shanghai 200240, China.

[2]State Key Laboratory of Micro-nano Engineering Science, School of Integrated Circuits (School of Information Science and Electronic Engineering), Shanghai Jiao Tong University; Shanghai 200240, China.

*Corresponding author: Weiwen Zou (wzou@sjtu.edu.cn), Shaofu Xu (s.xu@sjtu.edu.cn)
†These authors contributed equally to this work.

Supplementary notes S1-S6;
Supplementary figures S1-S6;
Supplementary references S1-S4.

**Supplementary notes**

**S1 Mutual information loss in the process of radar imaging**

As shown in Fig. S2a, during the radar-signal transceiver stage, given the radar-transmitted signal $s(t)$ and assuming that the TIR is $h(t)$, the received physical waveform $y(t)$ can be expressed as
$$y(t) = s(t) * h(t) + n(t)$$
To process the received physical waveform $y(t)$, the conventional method utilises analogue-to-digital conversion (ADC) to transform the analogue waveforms into a digital signal $y[n]$, and obtains the target image (high-resolution range profile (HRRP) or 2D image) $p[n]$ through imaging algorithms in the digital domain. Subsequently, the image features $f_1[n]$ were extracted, and classification and recognition were performed based on these features.

By directly processing electromagnetic waveforms, physical computing achieves higher recognition accuracy than radar imaging. This processing paradigm is based on DPI: $I(y;h) \geq I(p;h)$. For most radar imaging algorithms, equality does not hold (i.e., the imaging process inevitably causes mutual information loss). We use the BP algorithm for ISAR imaging as an example to discuss information loss during the imaging process. The BP algorithm primarily consists of three steps: (1) performing pulse compression on each physical waveform, (2) applying phase compensation to achieve zero phase for waveforms acquired at different times owing to variations in distances between points in the imaging area and the radar, and (3) coherently integrating the phase-compensated waveforms, where scatterer locations exhibit peaks due to coherent accumulation from the zero phase, thereby forming the target image (Fig. S2b).

According to the DPI, when processing is irreversible, equality does not hold, indicating mutual information loss. First, consider the pulse compression step in BP imaging, which involves the operation of matched filtering the waveform with the transmitted signal. For LFM transmission signals, the output of pulse compression can be represented as:

$$S_{ps}(t) = y(t) \otimes s^*(-t) = \sigma_P T_p \frac{\sin\left[\pi k T_p\left(1 - \frac{|t|}{T_p}\right)\right]}{\pi k T_p t} rect(\frac{t}{2T_p}) e^{i2\pi f_c t}$$

$$\approx T_p sinc\left[\pi k T_p\left(t - \frac{2R}{c}\right)\right] rect\left(\frac{t - \frac{2R}{c}}{2T_p}\right)$$

where $T_p$ denotes the pulse period, $B$ is the bandwidth, the frequency modulation slope is $k=B/T_p$, $rect(t/T)$ is the rectangular window function, and $R$ indicates the distance from the scattering point to the antenna.

This process is reversible because the transmitted signal is known. However, this reversibility presupposes the preservation of both amplitude and phase in the pulse compression results. While most existing radar HRRP recognition studies use only amplitude information[1,2], this step causes information loss (see main text, Fig. 2e, g).

In the phase compensation step, each waveform is multiplied by a compensation factor computed based on the positions of the grid points and the antenna. As long as these specific compensation factors are retained during computation, the process remains reversible. However, during coherent integration, all phase-compensated waveforms are summed to form a radar image.

$$I(i,j) = \sum_{m=1}^{M} S_{ps}(m, n_{ij}) e^{-j\frac{4\pi R_{i,j} f_c}{c}}$$

where $M$ denotes the total number of pulses required for imaging, $n_{ij}$ indicates the range cell corresponding to grid points $(i,j)$, and $e^{-j\frac{4\pi R_{i,j} f_c}{c}}$ represents the compensation factor.

This constitutes a many-to-one mapping, making it impossible to recover individual waveform amplitudes and phases from the integrated result, thus rendering this step irreversible. Finally, similar to HRRP recognition, ISAR image recognition more closely resembles image-processing tasks. Although ISAR images contain both magnitude and phase components, most studies retain only the magnitude information[3,4]. Therefore, in the ISAR imaging process, both coherent integration and output of ISAR images lead to information loss.

**S2 Comparison study of different processing methods**

Due to the limited interpretability of neural networks, the high classification accuracy achieved through physical computing alone does not provide sufficient evidence that acquiring and preserving more mutual information necessarily leads to higher recognition accuracy. Therefore, to validate our framework, we conducted a comparative study of different processing methods for various electromagnetic physical waveforms.

First, we employed LFM signals, commonly used in radar imaging, as the transmitted signals. The physical waveforms reflected from four target categories were fed into a convolution neural network for physical computing. The training results are presented in Fig. S3a, and the classification results are shown in Fig. 2e. The 89% recognition accuracy indicates that performing physical computation directly on the waveform effectively extracts target information and realises the sensing function.

Second, we investigated whether information loss occurs after pulse compression. This experiment consisted of two parts: processing both the magnitude and phase of the HRRP, and processing only the magnitude of the HRRP. These two cases correspond to the scenarios analysed theoretically in Supplementary note S1: the first case preserves mutual information, whereas the second loses phase information, resulting in mutual information loss. In the first case, the real and imaginary components were fed into the neural network as two distinct input channels. The network architecture was identical to that used for physical computing, except that the input of the first convolutional layer was modified to two channels. The training process is illustrated in Fig. S3b, and the classification results are shown in Fig. 2f. The comparable recognition accuracies of both methods (88% versus 89%) demonstrate effective information retention. In the second case, the magnitude of the HRRP was input into the neural network using the same architecture as that employed in the physical computing experiment. The training results are presented in Fig. S3c, and the classification results are shown in Fig. 2g. The recognition accuracy of 78% demonstrates that mutual information loss during the radar imaging process leads to significant accuracy degradation.

Finally, after obtaining the ISAR image of the target using the BP algorithm, we input the image into a 2D CNN for recognition. This CNN has a larger number of parameters than the neural network used in the physical computing experiment (Total parameters: 132.62 MB vs 11.45 MB). The training results are presented in Fig. S3d, and the classification results are shown in Fig. 2h. The recognition accuracy was only 75%, indicating that although radar images are more compatible with human visual perception, physically computing the original waveforms containing more mutual information can achieve better results for powerful feature extraction networks.

Based on the results of the comparative study, we confirmed that directly processing waveforms through physical computing is the fundamental reason for achieving high classification accuracy. Therefore, the validity of the concept of preserving mutual information is reinforced.

**S3 Theoretical derivation of the influence of autocorrelation function on mutual information**

Assuming that both the TIR and noise follow a circularly symmetric complex Gaussian distribution, the mutual information between the received signal and the TIR can be expressed as

$$I(y;h) = log|C_n^{-1}SC_hS^H + I|$$

where $C_n \in \mathbb{C}^{N \times N}$ represents the covariance matrix of noise, $S \in \mathbb{C}^{N \times N_h}$ corresponds to the transmit waveform convolution matrix, and $C_h \in \mathbb{C}^{N_h \times N_h}$ represents the covariance matrix of the TIR.

We assume that $C_n = \sigma^2 I$ (indicating that noise is dominated by the white noise) and $C_h = diag(\lambda_1, \lambda_2, \ldots, \lambda_{N_h})$ (indicating that the uncertainty of each element of the target impulse response vector is independent). Subsequently, the mutual information can be rewritten as

$$I(y;h) = log\left(|C_h^{-1}||\sigma^{-2}S^HS + C_h^{-1}|\right)$$

Under conditions where the target and the noise are fixed, to maximise mutual information, one can adjust the transmit waveform convolution matrix $S$ to increase $|\sigma^{-2}S^HS + C_h^{-1}|$. Note that:

$$S^HS = e_t \begin{pmatrix} 1 & r_1^* & \cdots & r_{N_h-1}^* \\ r_1 & 1 & \cdots & r_{N_h-2}^* \\ \vdots & & \ddots & \vdots \\ r_{N_h-1} & r_{N_h-2} & \cdots & 1 \end{pmatrix}$$

where $r_i = \frac{1}{e_t}\sum_{k=1}^{N_0-N_h} s_k^* s_{k+N_h}$ is the aperiodic autocorrelation function of $s$. $N_0$ is the length of $s$. $e_t = s^H s$ is the transmission power.

Let $M = \sigma^{-2}S^HS + C_h^{-1} = \sigma^{-2}e_tI + C_h^{-1} + \sigma^{-2}e_tR$, where $I$ is the identity matrix, $R$ is the off-diagonal matrix of $S^HS$, regarded as a perturbation term.

Let $D = \sigma^{-2}e_tI + C_h^{-1} = diag(\sigma^{-2}e_t + \lambda_i^{-1})_{1 \leq i \leq N_h}$, $R$ is redefined as $\sigma^{-2}e_tR$. According to the identity $|M| = exp(tr(\ln(M)))$, where $tr$ denotes the matrix trace, a Taylor expansion is performed on $M = D + R$:

$$\ln(D + R) = \ln D + \ln(I + D^{-1}R)$$

For sufficiently small $|r_i|$ values, the condition $\|D^{-1}R\| < 1$ is satisfied, ensuring convergence of the series expansion. Then: $\ln(I + D^{-1}R) = \sum_{k=1}^{\infty}(-1)^{k+1}\frac{(D^{-1}R)^k}{k}$

Thus:

$$|M| = |D + R| = |D| \cdot \exp\left(tr\left(\sum_{k=1}^{\infty}(-1)^{k+1}\frac{(D^{-1}R)^k}{k}\right)\right)$$

$$= |D|\exp\left(tr(D^{-1}R) - \frac{1}{2}tr((D^{-1}R)^2) + O(|r_i|^3)\right)$$

Since $D$ is a diagonal matrix and $R$ has zero diagonal elements and is conjugate symmetric, we have $tr(D^{-1}R) = 0$. The second-order terms are expanded as follows:

$$tr((D^{-1}R)^2) = \sum_{i,j}^{N_h} D_{ii}^{-1}D_{jj}^{-1}R_{ij}R_{ji} = \sum_{i\neq j}D_{ii}^{-1}D_{jj}^{-1}|R_{ij}|^2 = \sum_{i\neq j}\frac{\sigma^{-4}e_t^2|r_{|i-j|}|^2}{(\sigma^{-2}e_t + \lambda_i^{-1})(\sigma^{-2}e_t + \lambda_j^{-1})}$$

We can get:

$$|M| = \left(\prod_{i=1}^{N_h}(\sigma^{-2}e_t + \lambda_i^{-1})\right)\left(1 - \frac{\sigma^{-4}e_t^2}{2}\sum_{i\neq j}\frac{|r_{|i-j|}|^2}{(\sigma^{-2}e_t + \lambda_i^{-1})(\sigma^{-2}e_t + \lambda_j^{-1})} + O(|r_i|^3)\right)$$

Therefore, within the neighbourhood of $r_i = 0$, $(1 \leq i \leq N_h)$, a decrease in $|r_i|$ leads to an increase in $|M|$, thus the mutual information $I(y;h)$ also increases.

Through numerical simulations, we analysed the autocorrelation functions of various transmission signals. The Costas-LFM, generated by applying Costas coding to the carrier frequency of linear frequency modulated (LFM) signals, was selected as the desired transmit waveform using a Costas coding sequence of [2, 4, 8, 5, 10, 9, 7, 3, 6, 1]. As shown in Fig. 3c, compared with the LFM signals, Costas-LFM exhibits lower autocorrelation side lobes. We set $C_h = 0.1I$. Regardless of the noise power, the mutual information of Costas-LFM is always higher than that of LFM (Fig. 3c).

## S4 Information enhancement by Costas-LFM's HRRP

Figures. S4a and S4b show the confusion matrix and t-SNE results obtained after processing and recognising the intensity images formed from Costas-LFM physical waveforms in the HRRP. From the confusion matrix results, it can be observed that the recognition accuracy decreases slightly but still achieves relatively good performance. However, the t-SNE feature map shows that the features of different classes become closer (J16 and J20, Plane and UAV), indicating that after the imaging operation, there is indeed a loss of mutual information that makes different data features more similar and harder to distinguish. These results indicate that when receiving physical waveforms of the Costas-LFM, more mutual information is obtained compared to the LFM. Even though imaging operations cause information loss, the retained mutual information remains sufficient to ensure correct recognition.

## S5 Physical computing performance under different SNR conditions

The information extraction capability of physical computing in noisy environments was evaluated by introducing additive white Gaussian noise (AWGN) with varying power levels into the acquired Costas-LFM waveforms. The originally received physical waveforms had an initial SNR of approximately 16 dB, with noise injection expanding the SNR range from -15 dB to 15 dB. With 5 dB increments between adjacent SNR levels, this procedure produced seven distinct SNR conditions. These datasets were processed using both physical computing and HRRP recognition (with pulse compression), with the corresponding training trajectories and classification results for each SNR condition illustrated in Fig. S5.

**S6 Time complexity analysis**

We compared the time complexities of different processing methods: physical computing, HRRP recognition, and ISAR image recognition. Here, we assume that the length of the received physical waveform is $N$, and the detailed algorithmic steps and corresponding multiply-accumulate operations (MACs) are shown in Fig. S6.

In physical computing, the time complexity primarily includes two convolutional layers and two fully connected layers. In the 1D convolution computation, with a kernel size of three, a stride of two, and four output channels per layer, the MACs for the convolutional layers are $1 \times \frac{N}{2} \times 3 \times 4 + 4 \times \frac{N}{4} \times 3 \times 4$. After two convolutional layers with a stride of 2 and one pooling layer with a pool size of 8, the data volume input to the fully connected layer is $\frac{N}{32} \times 4 = \frac{N}{8}$, where 4 indicates that the convolutional layer output contains four channels. The first fully connected layer contained 300 neurons, and the second contained 4. Therefore, the total number of MACs for direct processing by electronics is $(1 \times \frac{N}{2} \times 3 \times 4 + 4 \times \frac{N}{4} \times 3 \times 4) + (\frac{N}{8} \times 300 + 300 \times 4)$ with a time complexity of $O(N)$.

For HRRP recognition, the initial step involves pulse compression on the received waveforms. Using a frequency-domain implementation approach, the required number of MACs is $2N\log N + N$. Therefore, the total number of MACs is $(2N\log N + N) + 1 \times \frac{N}{2} \times 3 \times 4 + 4 \times \frac{N}{4} \times 3 \times 4) + (\frac{N}{8} \times 300 + 300 \times 4)$.

During ISAR imaging, compensation factors are calculated according to the instantaneous relative geometry, followed by phase compensation of the pulse-compressed data. The total number of MACs required is $4N^2$. Each ISAR image requires coherent integration of M waveforms; therefore, the total MACs required for ISAR imaging are $M(2N\log N + N + 4N^2)$. For an ISAR image of size $N \times N$, the required MACs for the three-layer 2D convolutional network corresponding to the neural network parameters are $\frac{N^2 \times 3^2 \times 48}{2^2} + \frac{48 \times \frac{N^2}{2^2} \times 3^2 \times 128}{2^2} + \frac{128 \times \frac{N^2}{2^4} \times 3^2 \times 128}{2^2}$. The MACs required for the fully connected layers are $128 \times \frac{N^2}{2^6} \times 300 + 300 \times 4$. Consequently, the overall computational complexity of ISAR image recognition reaches $O(N^3)$, which substantially exceeds the computational demands of physical computing.

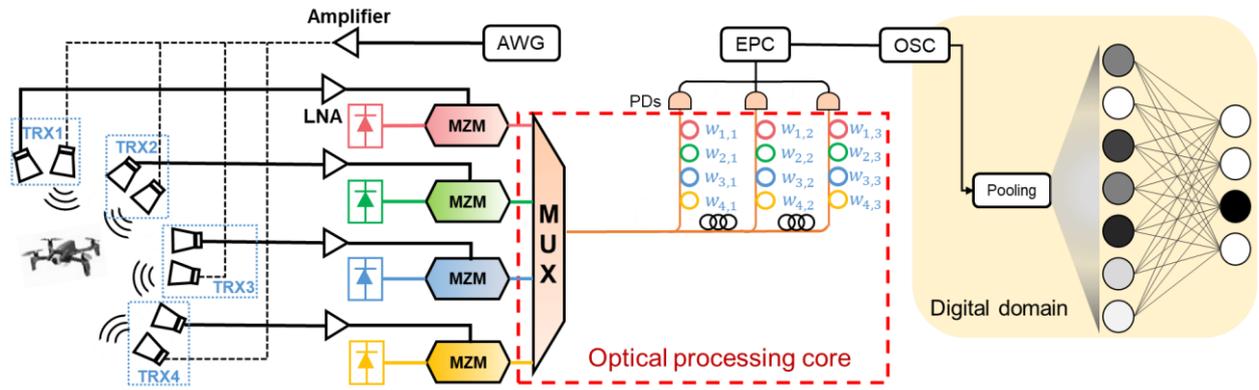

**Figure S1 | Experimental setup of PPCP.** Nonlinear and MaxPooling operations are implemented in the digital domain, with the fully connected layer outputting the recognition results.

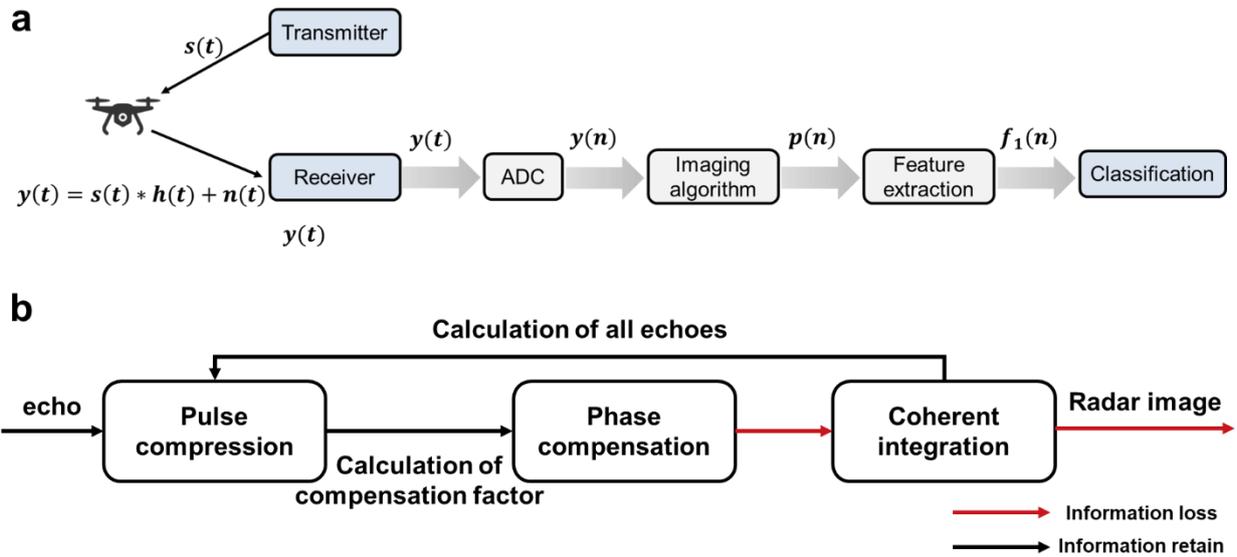

**Figure S2 | Information loss during BP ISAR. a**, Schematic of radar transceiver and signal processing. After the receiver receives the target physical waveform $y(t)$, the conventional method involves ADC sampling of the physical waveform, reconstructing the radar image of the target using radar imaging algorithms, and extracting image features for classification. **b,** Analysis of the back-projection algorithm in ISAR imaging demonstrates that both the coherent integration process and the radar intensity image formation stage inevitably cause information loss.

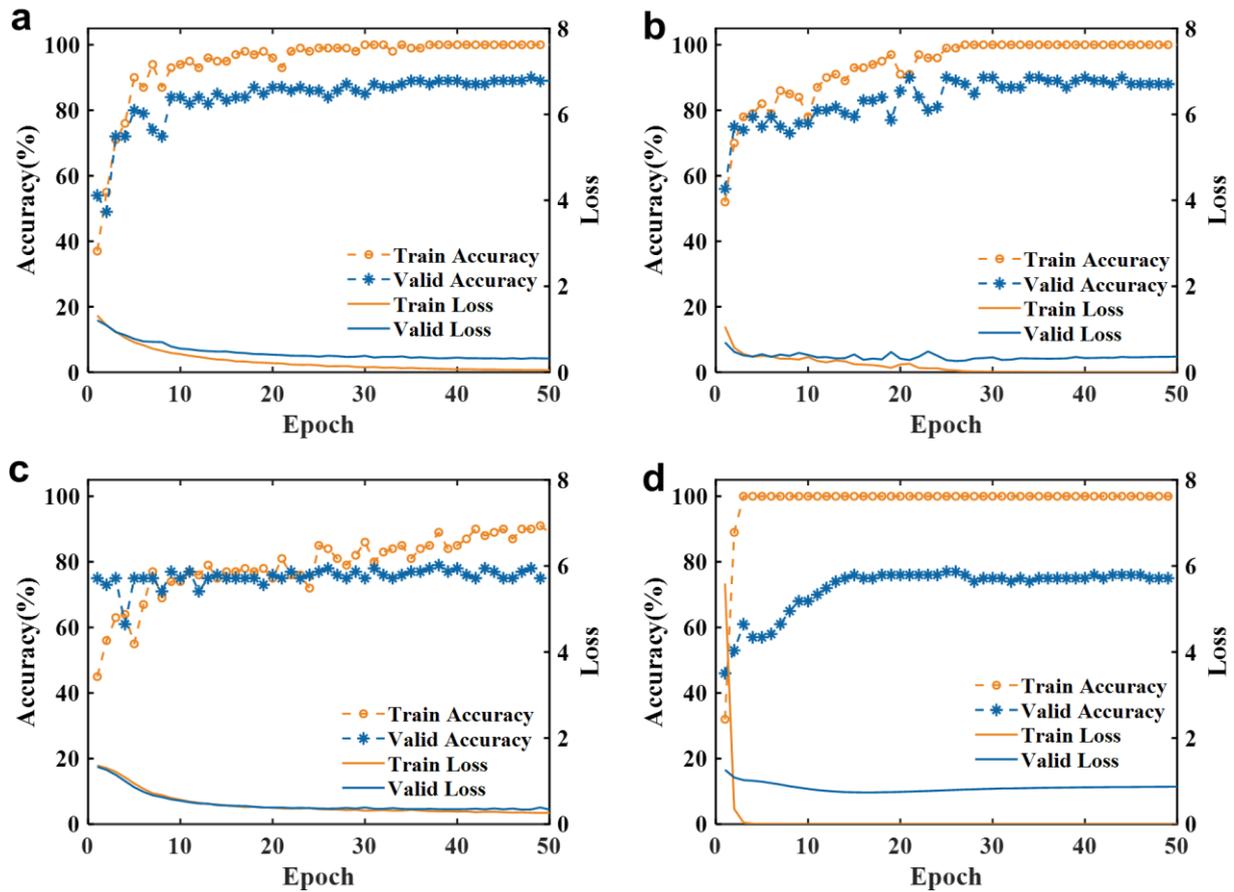

**Figure S3 | Training curves of the comparison study. a,** In the case where the input data are LFM physical waveforms, corresponding to the situation of using neural networks to directly process electromagnetic physical waveforms for recognition. **b,** In the case where the input data are the real and imaginary parts of the LFM waveform's HRRP, corresponding to the situation of performing pulse compression on physical waveforms to obtain target HRRP, then inputting both amplitude and phase information into neural networks for recognition. **c,** In the case where the input data are the amplitude components of the LFM waveform's HRRP, corresponding to the situation of using only amplitude information of the HRRP for recognition. **d,** In the case where the input data are ISAR images, corresponding to the situation of using the BP algorithm to construct ISAR images from physical waveforms, then inputting image information into neural networks for recognition.

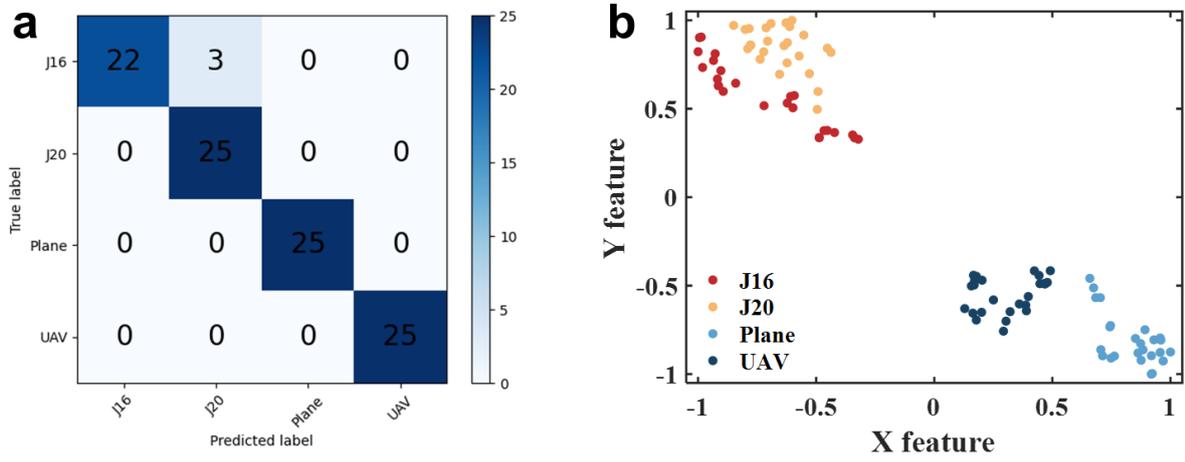

**Figure S4 | Recognition results of the HRRP obtained from transmitting Costas-LFM. a,** Confusion matrices for classifying the HRRP. **b,** Result of t-SNE when directly processing the physical waveforms.

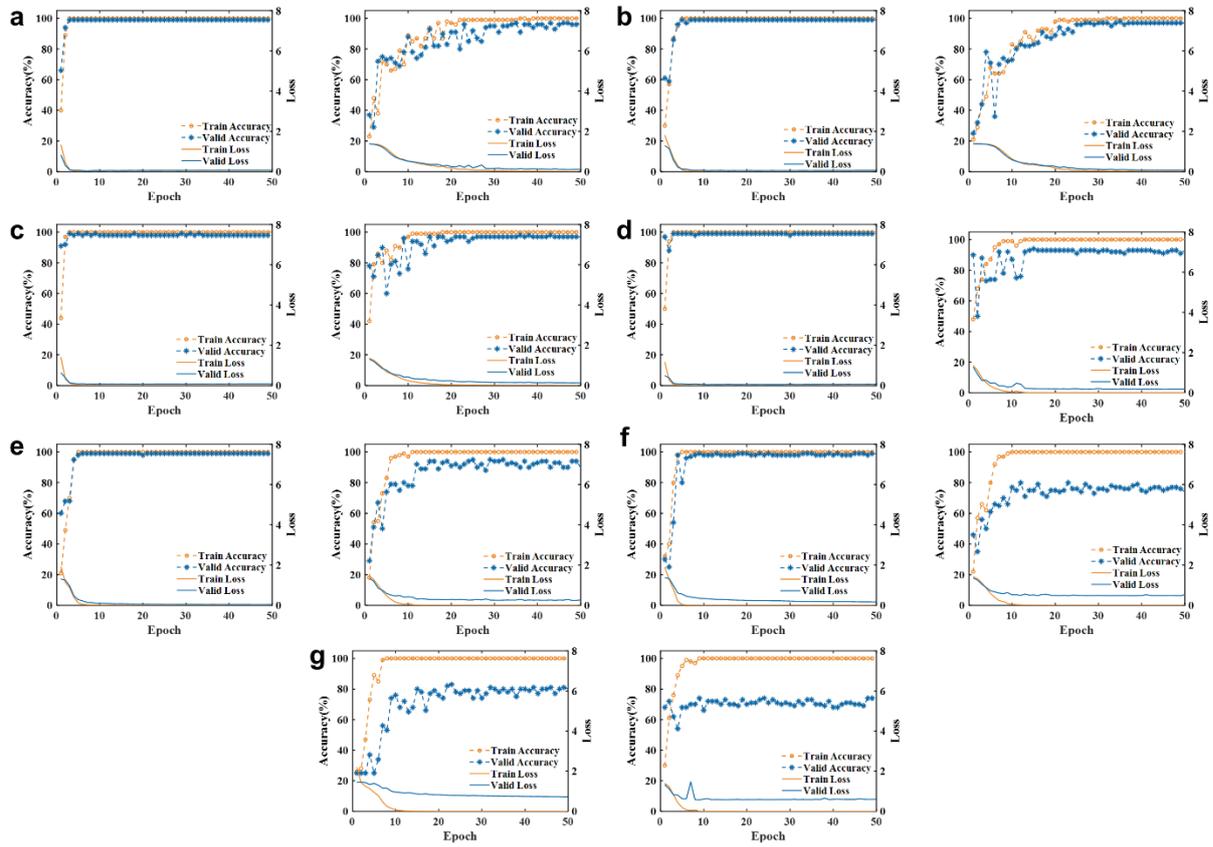

**Figure S5 | Training curves of the comparison study under different noise levels**. Training curves of physical computing (left) and HRRP recognition (right), respectively. **a,** 15 dB. **b,** 10 dB. **c,** 5 dB. **d,** 0 dB. **e,** -5 dB. **f,** -10 dB. **g,** -15 dB.

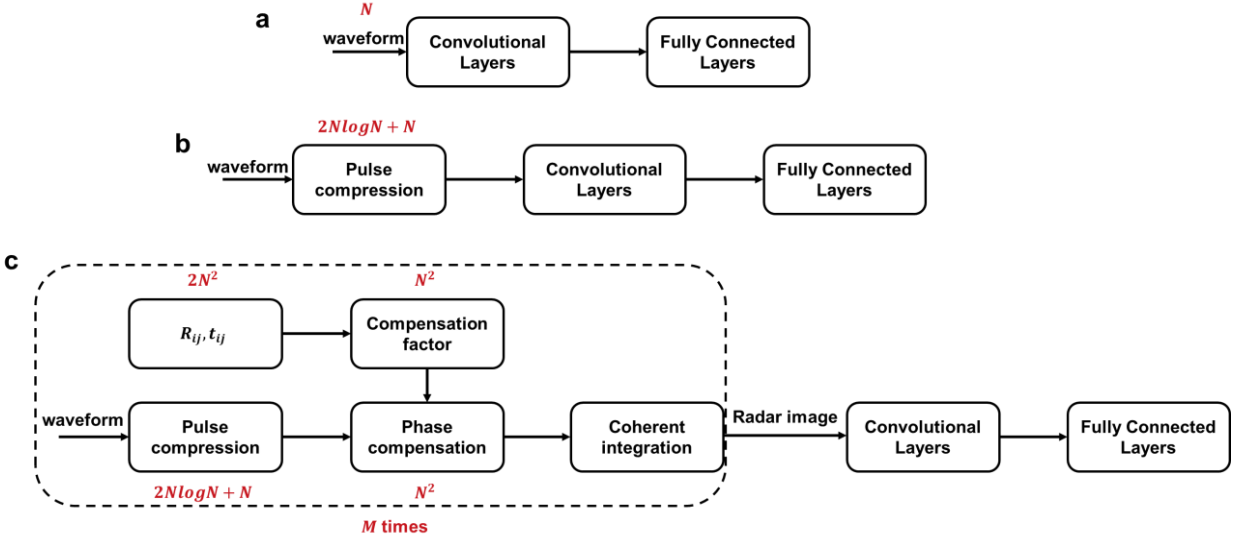

**Figure S6 | Time complexity of different processing methods**. **a,** Time complexity of physical computing. Performing convolutional and fully connected operations primarily contributes to the time complexity, with the overall system time complexity being $O(N)$. **b,** Time complexity of HRRP recognition. To obtain the HRRP of the physical waveform, pulse compression is required, resulting in a time complexity of $O(NlogN)$. **c,** Time complexity of ISAR image recognition. In the BP algorithm, in addition to the pulse compression operation, it is necessary to compute the distance and time delay from each imaging point to the antenna based on the imaging area, and calculate compensation factors to achieve phase compensation. The time complexity of this part is $O(N^2)$. As the BP algorithm requires the integration of M waveforms to form the target image, the total time complexity becomes $O(MN^2)$. Furthermore, in the ISAR image recognition process, because the input is a two-dimensional image, the time complexity of the recognition part is also $O(N^2)$.